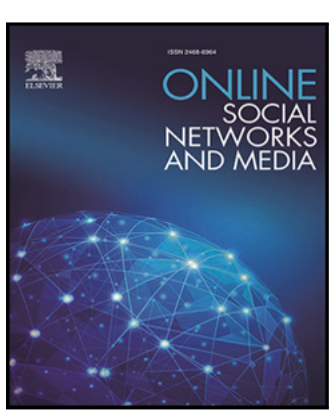

# IMMENSE: Inductive Multi-perspective User Classification in Social Networks☆

Francesco Benedetti [a,b,c], Antonio Pellicani [a,c,*], Gianvito Pio [a,c], Michelangelo Ceci [a,c,d]

[a] Department of Computer Science, University of Bari, Via Orabona 4, Bari, 70125, Italy
[b] Department of Computer Science, University of Pisa, Largo Bruno Pontecorvo 3, Pisa, 56127, Italy
[c] Data Science Laboratory, National Interuniversity Consortium for Informatics (CINI), Via Volturno 58, Roma, 00185, Italy
[d] Department of Knowledge Technologies, Jožef Stefan Institute, Jamova 39, Ljubljana, 1000, Slovenia

ABSTRACT

Online social networks increasingly expose people to users who propagate discriminatory, hateful, and violent content. Young users, in particular, are vulnerable to exposure to such content, which can have harmful psychological and social repercussions. Given the massive scale of today's social networks, in terms of both published content and number of users, there is an urgent need for effective systems to aid Law Enforcement Agencies (LEAs) in identifying and addressing users that disseminate malicious content. In this work we introduce IMMENSE, a machine learning-based method for detecting malicious social network users. Our approach adopts a hybrid classification strategy that integrates three perspectives: the semantics of the users' published content, their social relationships and their spatial information. Such contextual perspectives potentially enhance classification performance beyond text-only analysis. Importantly, IMMENSE employs an inductive learning approach, enabling it to classify previously unseen users or entire new networks without the need for costly and time-consuming model retraining procedures. Experiments carried out on a real-world Twitter/X dataset showed the superiority of IMMENSE against five state of the art competitors, confirming the benefits of its hybrid approach for effective deployment in social network monitoring systems.

## 1. Introduction

The significance of social networks in modern society is widely recognized as they enable various activities such as communication, networking, collaboration, and the sharing of news or interests. In addition, they also enable to perform several kinds of analysis and solve multiple tasks, such as user classification [1,2], profiling [3,4], and recommendation [5–8].

Over time, social media platforms have undergone substantial evolution, allowing users to form multiple types of relationships, for instance, by *liking* posts, *sharing* others' opinions, *engaging* in real-time conversations, and *joining* groups or communities. Nevertheless, these platforms can also be misused for malicious purposes, including *(i)* disseminating fake news, promoting hate against minorities, and spreading radical ideologies, *(ii)* assembling like-minded individuals into extremist communities, and *(iii)* recruiting vulnerable individuals into terrorist or criminal organizations. In this regard, several reports have highlighted how radicals exploit social media platforms [9,10].

To counter this phenomenon, Law Enforcement Agencies (LEAs) established dedicated divisions whose job is to detect and take action against users who engage in illegal activities. However, social networks are dynamic systems in continuous evolution, generating a massive amount of data that humans cannot monitor in real time. In this scenario, it has become essential to assist LEAs with automated tools capable of detecting inappropriate content and reporting suspicious users. Specifically, for this task, three main dimensions should be taken into account: *(i)* the content generated by users, since it can reveal beliefs and intentions; *(ii)* the social relationships established with other users and content thereof (i.e., *follows*, *retweets*, *likes*), to identify silent users who do not post inappropriate content but are in touch with other users appearing as suspicious; *(iii)* the spatial relationships, possibly established by geographical closeness among users, since dangerous nearby users are more willing to join local communities. These three dimensions, also referred to as *modalities*, are usually treated by existing approaches for user classification in a separate manner. In particular,

☆ This article is part of a Special issue entitled: 'disinformation-toxicity-harms' published in Online Social Networks and Media.

* Corresponding author at: Department of Computer Science, University of Bari, Via Orabona 4, Bari, 70125, Italy.
*E-mail addresses:* francesco.benedetti@uniba.it (F. Benedetti), antonio.pellicani@uniba.it (A. Pellicani), gianvito.pio@uniba.it (G. Pio), michelangelo.ceci@uniba.it (M. Ceci).

most of the existing systems perform the classification based on the content published by users [11,12], or solely based on the topology of the network established based on social or spatial relationships [13,14]. However, online radicalization is a complex phenomenon that involves several aspects. Considering only one of them and disregarding the others can result in not detecting users who are dangerous under a different, unconsidered perspective.

For this reason, hybrid strategies that incorporate a combination of different modalities for analyzing social networks have been proposed. These strategies have been successfully used in various user classification tasks in social networks, showing improved performance compared with tools that use a single modality/perspective. Among them, focusing on methods aiming to identify suspicious users, that is the goal of the method proposed in the present paper, the method proposed in [1], called SAIRUS, can be considered the first approach that simultaneously leverages the three aspects mentioned above in a multimodal fashion. However, SAIRUS is limited in the approach adopted to represent nodes, as it operates under a *transductive* [15] setting. In this setting, a classification model is learned from a network containing both labeled and unlabeled nodes, and is then exploited to assign labels to the unlabeled nodes (see Fig. 1(a)). Thus, when training models in such a setting, nodes that need to be labeled have to be already available at training time. As a consequence, it is impossible to classify any node that has not been observed during the training phase. It is therefore noteworthy that a transductive approach cannot be easily adopted in a real-world scenario, since LEAs would need to retrain the model every time a new user needs to be assessed.

To overcome this limitation, in this paper we propose IMMENSE (**I**nductive **M**ulti-perspective **M**odel for us**E**r classificatio**N** in **S**ocial n**E**tworks), a new multimodal approach that works in the *inductive* setting. In particular, IMMENSE learns a model from a labeled network of users provided at training time, that is able to generalize to new, unseen nodes by leveraging the different modalities considered (see Fig. 1(b)). To the best of our knowledge, no existing technique is able to solve the suspicious user identification task in such a setting, simultaneously leveraging the above-mentioned modalities/perspectives (content, social relationships, and spatial relationships). Therefore, the main contributions of this paper can be summarized as follows:

- we propose the novel method IMMENSE that can be considered among the first approaches that consider multiple perspectives, i.e., the posted content, the social relationships, and the spatial relationships, for the identification of risky users in social networks;
- methodologically, IMMENSE is able to work in the inductive setting, enabling the learned model to be directly adopted for making predictions on new users who have not been observed during the training;
- as a consequence, IMMENSE promotes model reusability and, therefore, the sustainability of AI systems, since, in real-world environments, making predictions for a batch of new users does not require to re-train a model from scratch.

The remaining of the paper is organized as follows: in Section 2 we briefly discuss some related work; in Section 3 we describe the details of the proposed approach; in Section 4 we describe the results of our experimental evaluation; finally, in Section 5 we draw some conclusions and outline possible future works.

## 2. Related work

Since this work has its roots in the node classification task for network data, in the following subsection we briefly describe some of the existing transductive and inductive techniques for solving such a task. Then, we provide an overview of existing approaches for user classification in social networks.

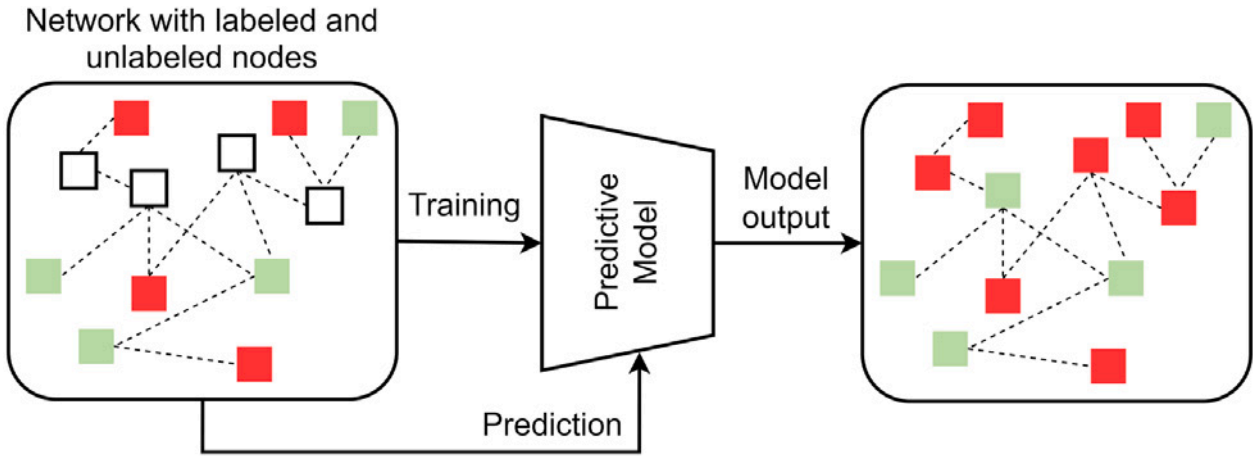

(a) Transductive learning setting

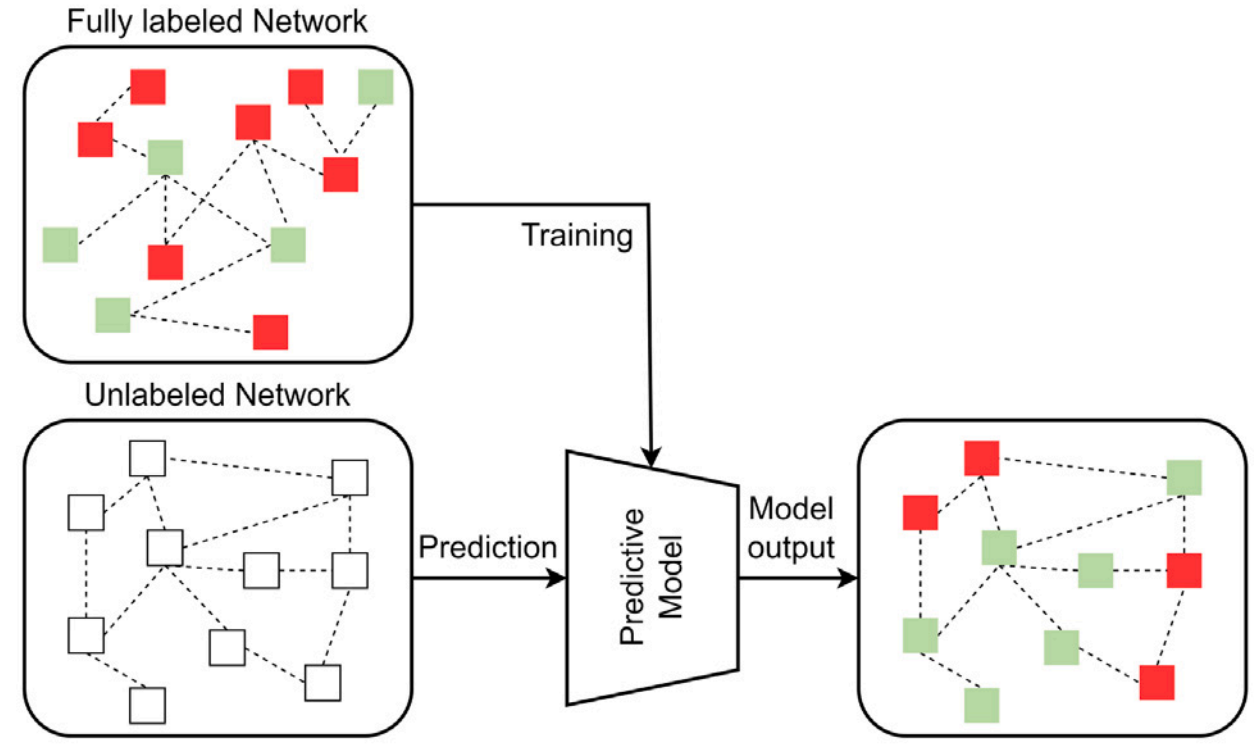

(b) Inductive learning setting

**Fig. 1.** A graphical representation of the transductive (on the top) and inductive (on the bottom) learning settings. White nodes represent unlabeled nodes.

### 2.1. Transductive and inductive learning for network data

Early techniques for solving machine learning tasks from network data rely on identifying a numerical representation of nodes (node embeddings) in a transductive setting [16–20], and learning a predictive model on top of such representations. These techniques are well-suited for scenarios where the network is only partially labeled and the considered task consists of labeling observed nodes only. In this context, pioneering methods for learning node embeddings are DeepWalk [21] and Node2Vec [22], both based on random walks. Nevertheless, we can also find methods working in the transductive setting that directly classify unlabeled nodes without performing a preliminary embedding phase. An example is GNetMine [13], which works with heterogeneous networks, performing label propagation across nodes of different types.

On the other hand, techniques able to identify node embeddings in an inductive setting aim to learn a decision function across the entire data space, thus enabling the generalization to unseen nodes that possess similar types of features. One of the most popular approaches is GraphSAGE [23], which operates in an iterative fashion. In particular, it aggregates the features of the nodes in the immediate neighborhood of the node being represented: At each iteration, the representation of a given node at the previous iteration is concatenated with that obtained through the aggregation of the representations of its neighboring nodes, and processed by a nonlinear activation function to compute the new representation. In the inference phase, the learned model is applied to represent a new (unseen) node, also aggregating information from its local neighborhood.

More recently, the method Heterogeneous Graph Transformer [24] (HGT) has been specifically proposed for the analysis of heterogeneous graphs. To model heterogeneity, HGT employs multiple parameters that are node-type and edge-type dependent to characterize the heterogeneous attention over each edge, empowering it to maintain dedicated representations for different types of nodes and edges.

Another method that works in the inductive learning setting is AGAIN [25]. It combines a sampling approach with an attention-based strategy. During the sampling phase, a fixed number of neighbors are selected, with the size varying according to the search depth. Then, during the aggregation, an attention mechanism assigns different learnable weights to the neighbors, indicating their relative importance for the embedding of a given node.

Finally, it is worth mentioning RIO-GNN [26], an inductive node embedding framework for multi-relational graphs. Its architecture consists of three modules: the first measures the similarity among training instances sharing the same label; the second uses these similarity values to sample, for each instance, a set of closest neighbors under each relationship; the third aggregates neighbor information across relations to obtain a comprehensive embedding.

In the literature, we can also find approaches capable of operating in both inductive and transductive settings. An example is represented by the popular Graph Attention Networks (GATs) [27], which calculate node representations using multi-head attention [28]. In the transductive setting, GATs process the entire network simultaneously, learning attention weights that model the relative importance of node relationships in the existing graph structure. However, GATs do not require that all nodes are available during training, as their shared attention mechanisms can generalize to unseen nodes. When new nodes become available, the learned attention functions are applied inductively to process their neighborhood structures and properties.

Another approach capable of operating in both settings is Graph Extrapolation Network [29], having the goal to perform Out Of Graph (OOG) link prediction. The representation for a new node is computed by exploiting the features of its neighboring nodes and edges. At training time, in order to incorporate knowledge about the relationships between unseen nodes, a transductive component is employed. This involves aggregating the representations of neighboring nodes, also considering unseen nodes, which are generated/simulated at training time via a meta-learning framework.

### *2.2. User classification in social networks*

The user classification task aims at predicting a label for unlabeled users in a network, based on their properties. The labels may vary depending on the specific considered problem. Existing user classification techniques fall into three categories, namely: content-based, topology-based, and hybrid.

**Content-based** approaches classify users by relying exclusively on their published content.

For example, in [2] the authors aim at detecting bot accounts. Users are represented as a document obtained by concatenating their published posts. Documents are divided into a fixed set of chunks and, for each term in the vocabulary, its occurrences in every chunk are counted, obtaining a matrix for each document, where the rows represent chunks and columns represent the number of occurrences in a chunk. A weighting scheme is then used to adjust the term signals to better capture differences between human and bot-generated texts. The signals are decomposed via wavelet transform functions and, lastly, a set of features is extracted from the transformed matrix and fed to a classifier (authors experimented with MLPs and random forests) for learning a binary classification model.

A content-based user classification approach was also used for predicting the political affiliation of Twitter users [30]. The authors adopted a two-step approach: first, they use a model based on Bi-LSTM and attention, trained on a corpus of tweets wrote by members of the U.S. Congress, for computing the probability of a single tweet to be associated with the democratic or the republican party. Then, for each user, they average the probabilities computed for their tweets and compute additional statistics, such as standard deviation, minimum, maximum, median, first quartile, third quartile, proportion of tweets with a high probability (>0.6) of being Republican/Democratic. A Support Vector Machine is then used to predict the user's political affiliation based on these features.

Shifting to focus on **topology-based** classification approaches, it is worth noting that they aim to solve the learning task by performing exclusively the analysis of the topology of the network of relationships. State-of-the-art approaches are mainly based on graph neural networks (GNN) [31], which leverage the properties of the graph to create meaningful representations of nodes (users, in this case), edges (relationships among users), and even the whole graph (the whole social network). In [32] the authors purposely ignored node features to rely solely on the network topology, aiming to show the importance of node connections for user classification. They compared the performance of a model based on harmonic functions [33] with two GNN variants, namely, Graph Convolutional Networks (GCNs) [34] and Graph Attention Networks (GATs) [27], where the value of node features is purposely set to zero. All three classifiers achieved good results in terms of accuracy, with the harmonic classifier slightly outperforming the two competitors, possibly because natively based on network connections instead of node features.

Topology-based approaches have also been investigated for the exploitation of the spatial information. An example is [35], where the goal is to detect areas at higher risk for dengue. The authors analyzed tweets published by users who mentioned having personal experience with the disease. Clustering was performed using the geolocations associated with the tweets, followed by the application of two probabilistic models to identify spatial regions with a higher risk of infection.

Finally, **hybrid approaches** aim to combine the contribution provided by the content posted by users with that of the network topology.

An example of hybrid approach is Gitsec [36], a classification system for Github users, aimed at detecting malicious accounts on the platform. The framework solves classification tasks by taking into account the descriptive features about the accounts, their activity, and two networks: a user-repository graph, that tracks the way each user interacts with repositories, and a user-user graph, that describes the interactions among users. Methodologically, Gitsec uses two separate PLSTM networks connected with the attention mechanism to analyze data about the user activity. A GNN model based on GraphSAGE produces a prediction for the users based on the user-repo graph. The user-user graph is also analyzed and five structural features are obtained for each user. The outputs of these modules, together with the descriptive features of a user, are then fed to a decision maker, such as XGboost, to provide the final prediction for the users.

In [37], the authors propose a hybrid user classification framework that considers profile features, user generated content, social relationships and social interactions. The proposed system is made of a static module, that extracts statistical features from user profiles and user generated content. A second parallel module based on LSTM extracts dynamic features from user interactions. The two features sets are fed to an XGBoost based classifier to provide the final prediction.

Hybrid approaches can also be in the form of general-purpose methods able to solve classification tasks in the relational setting, where each perspective is modeled through one or more tables of a relational database. A relevant example is the system Re3py [38], a classification system based on ensembles of relational decision trees. The method first generates complex aggregate features by navigating foreign key paths involving multiple database tables. Then, a predictive model is learned via top-down induction of decision tree ensembles, that exploit both the original features and the generated ones. In the same category, we can also find the system Mr-SBC [39], a probabilistic approach that extends the naïve Bayes classifier to handle multirelational settings represented as database tables connected by foreign key constraints. The posterior probabilities are computed using first-order classification rules that are learned during the training phase.

It is noteworthy that most of the existing works primarily consider the content posted by users. While some approaches consider the relationships, they often create artificial links based on similarity measures or the co-occurrence of words, rather than explicitly modeling real relationships among users. Moreover, possible spatial relationships among users are almost always ignored. Finally, general-purpose relational methods, like those mentioned before [38,39], do not explicitly capture the semantics of the posted content, but represent it only through a simple document-word table of a relational database.

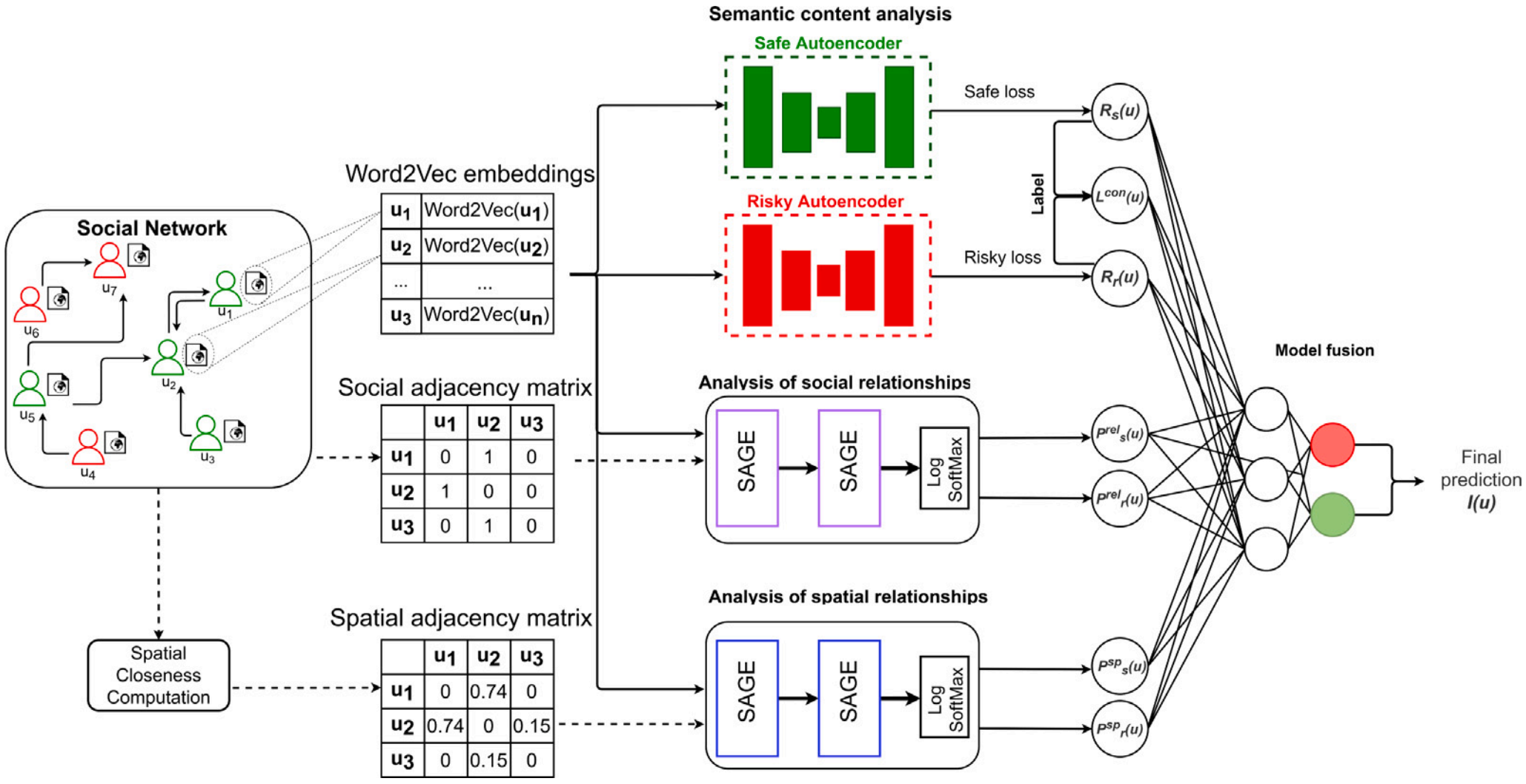

**Fig. 2.** Schema of the IMMENSE framework.

To the best of our knowledge, only one existing method [1] fully integrates the three aspects considered in this paper, also providing the right importance to the semantic of the content posted by users. However, as introduced in Section 1, it exhibits the limitation of transductive approaches, i.e., it cannot perform predictions for new users that have not been observed during the training phase, without re-training the whole model from scratch. On the contrary, the method IMMENSE proposed in this paper fully exploits these three perspective in an inductive learning setting.

## 3. The proposed method IMMENSE

Before describing the proposed method, we formalize some key aspects of the task we intend to solve. Specifically, we define a social network as $\langle N, C, E_C, E_T \rangle$, where:

- $N$ is the set of nodes, each representing a user. Users can be labeled as *safe* ($S$) or *risky* ($R$). We refer to the former as $N^{(S)}$ and to the latter as $N^{(R)}$. Some users may not be labeled (i.e., their label is unknown), meaning that $N^{(S)} \cup N^{(R)} \subseteq N$.
- $C$ is the set of textual contents posted by users. The geographical position from which the content was posted may be available.
- $E_C \subseteq N \times C$ is the set of relationships that link the users to the textual content they posted.
- $E_T \subseteq N \times N$ is the set of topological social connections among users. These relationships are represented by directed links, since are not necessarily symmetric (e.g., if user $A$ *follows* user $B$, it does not imply that user $B$ *follows* user $A$).

Since we work in a supervised inductive setting, the training phase exclusively involves labeled users. By leveraging their content, their social relationships and their geographical proximity, we train a model to accurately map users to their respective labels, ensuring it can generalize to new, unseen users.

The general architecture of the proposed method IMMENSE is depicted in Fig. 2, which shows its three main components, specialized for the analysis of the three considered perspectives. A final model fusion step is then performed to combine the contribution of such components. In the following subsections we describe and discuss each phase in detail.

### 3.1. Semantic content analysis of the textual content

This phase analyzes the textual content to generate a user profile and to assign a label according to such a perspective. Specifically, we process the textual content produced by users on the social network, applying a preprocessing pipeline that includes tokenization, stopword removal, and stemming. The resulting posts are then concatenated into a single document for each user, following their chronological ordering. In this way, IMMENSE can also indirectly take into account the way in which the content posted by the user evolved over time. This is possible through the adoption of a Word2Vec model [40], which is trained to map each word, considering its context in the text, into a $k_c$-dimensional semantic space. By leveraging the additive compositionality property of word embeddings [41], we compute an embedding vector for each available user. This property allows the user's embedding to be represented as the sum of the embeddings of the words in their posted contents. Note that to perform this step, also more recent approaches based on BERT or large language models (LLMs) could be used, but we stick with Word2Vec as it demonstrated superior accuracy in previous studies [42,43]. Moreover, learning a Word2Vec embedding is considerably less computationally intensive than training models like BERT or LLMs.

Subsequently, we use the obtained embeddings to train two distinct one-class classifiers (one for each class), which are based on stacked autoencoders [44]. An autoencoder is a model that compresses the input into a lower-dimensional representation through an encoder and then reconstructs the original input using a decoder. Formally, given an input $X$, an autoencoder learns an encoding function $en : X \rightarrow X'$ and a decoding function $dec : X' \rightarrow X$ such that:

$$(en, dec) = argmin_{(en,dec)} \| X - dec(en(X)) \|^2$$

In other words, the goal is to minimize the reconstruction error, which indicates how much the actual input differs from the reconstruction. In our model, the encoder is made of the input layer followed by two linear layers whose dimension is respectively one half and one fourth of the input. The decoder's layers mirror the structure of the encoder, so that the output has the same dimension as the input. Each layer adopts the ReLU as non linear activation function.

We employ one autoencoder trained on the embeddings of users labeled as *risky* and one autoencoder trained on those labeled as *safe*. Each user's representation is fed into both autoencoders, and their respective reconstruction errors, denoted as $R_R$ and $R_S$, are computed as the mean squared error between the input and the output vectors. The semantic content analysis module outputs these two reconstruction error values, along with a label $L^{(con)}$. The label is assigned a value of 0 (*safe*) if $R_R < R_S$; otherwise, it is set to 1 (*risky*). The adoption of two separate autoencoders, rather than a single binary classifier, is motivated by the fact that they tend to provide a higher accuracy in presence of class imbalance [45], which is the situation we face in our scenario.

### 3.2. Network topology analysis of user social relationships

This component analyzes the relationships between users in the social network. Specifically, the user relationships can be represented by an adjacency matrix $A \in \{0,1\}^{|N|\times|N|}$, where $A_{ij} = 1$ if user $N_i$ follows[1] user $N_j$, 0 otherwise. However, this matrix would have an extremely high dimensionality with significant sparsity, making it impractical to process. This occurs because, in most social networks, each user is connected with only a small fraction of the total users in the network, resulting in many entries of the adjacency matrix being zero. To address this issue, before learning a predictive model, we aim to identify a lower-dimensional, denser, feature space, of dimension $k_r$, representing the network of relationships of each user. For this purpose, we adopt GraphSAGE [23], that generates embeddings for nodes in a network in an iterative fashion, also exploiting the features of the nodes coming from the local neighborhood.

Formally, at the $i$th iteration and for each user $v \in N$, its embedding $h_v^i$ is computed also considering its neighborhood $N(v)$. In particular, we first produce an aggregated vector representation $h^i_{\mathcal{N}(v)}$ of its neighborhood at the $i$th iteration as:

$$h^i_{\mathcal{N}(v)} = \frac{\sum_{u\in\mathcal{N}(v)} h_u{}^{i-1}}{|\mathcal{N}(v)|} \tag{1}$$

Then, the current step user representation $h_v^i$ is obtained by concatenating its previous representation $h_v^{i-1}$ to the aggregated neighborhood information $h^i_{\mathcal{N}(v)}$ and by feeding the obtained vector through a fully connected layer, whose weights are updated in the training phase, with nonlinear activation function $\phi$ (in our case, the *ReLU* function):

$$h_v^i = \phi\left(\mathbf{W}^i \cdot CONCAT\left(h_v{}^{i-1}, h^i_{\mathcal{N}(v)}\right)\right) \tag{2}$$

where $\mathbf{W}^i$ is a learnable weights matrix. The obtained user representation serves as input for subsequent iterations in the algorithm, each of which is performed by a GraphSAGE layer. Therefore, the number of iterations corresponds with the number of GraphSAGE layers used to generate the final node embedding, as well as to the depth in the network reached when representing each node: each additional GraphSAGE layer expands the node's informational reach by one hop, effectively broadening the considered neighborhood. This layered approach allows the algorithm to capture increasingly complex structural information from the graph, enhancing the richness of the resulting embeddings.

In IMMENSE we construct a graph where users are connected based on their relationships in the analyzed social network. Each node (user) in the graph is associated with features corresponding to the user's semantic representation obtained in the previous phase of the method. It is noteworthy that this choice may appear to possibly introduce some redundancies with respect to the semantic analysis module. However, it is noteworthy that users tend to connect with others who share similar views and beliefs, which are primarily expressed through contents posted on social media platforms. By integrating the features extracted from the content, we not only capture the structure of connections, but also the underlying reasons for these connections. This approach enables a more comprehensive understanding of the network, where topology and content mutually inform one another. Essentially, we argue that these features do not introduce substantial redundancy in the representation; instead, they enhance the topological analysis by incorporating semantic information that captures the real-world factors driving network formation.

In IMMENSE we train GraphSAGE in a supervised manner to incorporate user labels from this early stage of the embedding. This approach helps the model to build a representation that facilitates reliable classification in the subsequent steps. To this end, our training architecture consists of three layers, where the first two layers are devoted to capturing the features of neighboring nodes of up to hops in the network, while the final is a linear layer with log-softmax activation function to output the predicted log-probabilities of the node to belong to the safe and risky class. The prediction is used to compute the loss, which is then backpropagated during training.

Another important aspect to consider, already mentioned before, is that the task being tackled is inherently imbalanced. While it is difficult to quantify precisely how much content on social media reflects extremist or discriminatory ideologies, it is reasonable to assume that such content represents only a small minority of the overall material available online. Hence, during the learning phase, users cannot be treated equally, as the *safe* class significantly outnumbers the *risky* class. This imbalance could lead the model to develop a bias towards the majority class [46]. As a consequence, errors made during the predictions cannot be treated equally, because misclassifying a risky user as safe is much more dangerous than mistakenly predicting that a safe user is risky. To address class imbalance, we consider two alternative strategies:

- **Class Weighting.** This strategy consists in using a weighting schema based on inverse class frequency. In particular, it assigns a weight to each instance, which is inversely proportional to the number of instances belonging to the same class in the training set. More formally:

$$weight_{risky} = \frac{|N|}{|N^{(R)}|} \qquad weight_{safe} = \frac{|N|}{|N^{(S)}|} \tag{3}$$

As training loss function we employ the negative log likelihood. The weights are incorporated into the training loss to give a higher penalty to the model when, during training, it misclassifies instances belonging to the minority class. Formally, let $p_{i,y_i}$ be the predicted probability that the model assigns to the $i$th instance $x_i$ of belonging to the true class $y_i$, the loss is defined as:

$$\mathcal{L}(x_i) = -weight_{y_i} \cdot log(p_{i,y_i}) \tag{4}$$

where $weight_{y_i} = weight_{risky}$ if $y_i = 0$, otherwise $weight_{y_i} = weight_{safe}$.

- **Focal loss.** This strategy employs the focal loss [47], that is a training loss function that directly accounts for the imbalance. It is a variant of the standard cross entropy function, defined as:

$$\mathcal{L}(x_i, risky) = -\alpha_{risky}(1-\tilde{p}_i)^\gamma \cdot log(\tilde{p}_i) \tag{5}$$

where $\alpha_{risky}$ is a balancing weight, $\gamma \geq 0$ is the focusing tunable parameter, and $\tilde{p}_i$ is computed as:

$$\tilde{p}_i = \begin{cases} p_i & \text{if } y = risky \\ 1-p_i & \text{otherwise} \end{cases} \tag{6}$$

In practice, the focal loss penalizes wrong false negatives on which the model has high confidence.

Once trained, the node classifier is used to compute, for each user $u$ in the training set, two probability values, respectively $P_s^{rel}(u)$ and $P_r^{rel}(u)$, which are the predicted probabilities that the user $u$ has of being safe or risky, based only on its social connections in the graph. Such probability values are the output of the relational analysis module, and will be subsequently used in the model fusion phase.

[1] In this formalization, we use the *follows* relationship as an illustrative example. More generally, any asymmetric or symmetric relationship (the latter achieved by duplicating links) can be modeled in a similar way.

### 3.3. Analysis of the user spatial closeness

This module is devoted to the analysis of the spatial closeness among users in the social network. While we again use an adjacency matrix to represent spatial relationships, the approach slightly differs from that adopted in the previous module. In particular, we construct a matrix $S \in [0,1]^{|N|\times|N|}$, where, unlike the matrix $A$ containing binary values used for the analysis of user relationships, each cell $S_{ij}$ contains a continuous closeness score in $[0,1]$. This score quantifies the spatial proximity between users $i$ and $j$, providing a granular representation of their spatial relationships.

To compute the closeness score, we approximate the position of each user by identifying the most frequent location associated with his/her posted contents. We opt for the most frequent location rather than other possible aggregations of locations, such as the average latitude/longitude, as the latter could potentially generate coordinates corresponding to a place where the user has never actually been.

Using these locations, we then compute the geodetic distance between the users of the social network. Specifically, given two users $u_1, u_2$ with their latitudes $\phi_1, \phi_2$ and their longitudes $\lambda_1, \lambda_2$, the distance $d(u_1, u_2)$ is given by:

$$d(u_1,u_2) = 2r \cdot \arctan\left(\sqrt{\frac{a(u_1,u_2)}{1-a(u_1,u_2)}}\right) \tag{7}$$

where $r$ is the Earth radius (≈6371 km) and $a(u_1,u_2) = \sin^2\left(\frac{\phi_1-\phi_2}{2}\right) + \cos(\phi_1)\cdot\cos(\phi_2)\cdot\sin^2\left(\frac{\lambda_1-\lambda_2}{2}\right)$. To obtain a closeness score from such distance values, we need to compute the z-normalized distances $z(u_1,u_2)$. First we define the mean $\mu_d$ and the standard deviation $\sigma_d$ of the distances as:

$$\mu_d = \frac{1}{|N|} \cdot \sum_{a,b\in N, a\neq b} d(a,b) \qquad \sigma_d = \sqrt{\frac{\sum_{a,b\in N,a\neq b}(d(a,b)-\mu_d)^2}{|N|}} \tag{8}$$

Then, the z-score normalized distance among users $(u_1, u_2)$ is computed as:

$$z(u_1,u_2) = \frac{d(u_1,u_2)-\mu_d}{\sigma_d} \tag{9}$$

This allows us to distinguish between users who are closer than the average and those who are farther than the average. More formally, since two users $u_1, u_2$ are closer than the average if $z(u_1,u_2) < 0$, we compute the closeness score in $[0,1]$ as follows:

$$closeness(u_1,u_2) = \begin{cases} \frac{z(u_1,u_2)}{min_z} & \text{if } z(u_1,u_2) < 0 \\ 0 & \text{otherwise} \end{cases} \tag{10}$$

where $min_z$ represents the minimum (most negative) z-normalized distance observed across all user pairs in the network. This normalization ensures that the user pair with the smallest z-score (i.e., the geographically closest pair) receives a closeness score of 1, while pairs with positive z-scores (farther than average) receive a closeness score of 0.

The obtained spatial adjacency matrix is used to build a graph, where two users $u1, u2$ are connected when $closeness(u_1,u_2) > 0$, with a weight equal to $closeness(u_1,u_2)$. We employ GraphSAGE also in this case, to extract $k_s$ dimensional embeddings of nodes. Note that a given user who never shares his/her geographical information in any of his/her posted content will appear as isolated, namely, spatially distant from all the other users. In this case, the embedding identified through GraphSAGE for the spatial dimension will solely rely on his/her initial features, that, as detailed in Section 3.2, are based on the embedding of the posted content.

To account for the class imbalance, we adopt the same strategies as we defined for the network topology analysis (see Section 3.2).

This module outputs, for a given user $u$, the probabilities $P_s^{sp}(u)$ and $P_r^{sp}(u)$, that are the predicted probabilities that the user is safe or risky, respectively, based on spatial relationships.

**Table 1**
Quantitative information about the considered dataset.

| | |
|---|---|
| Users | 37 945 |
| Risky users | 2 807 |
| Safe users | 35 138 |
| Users with spatial information | 1 043 |
| Avg document length | 136 |
| Avg social following per user | 16 |
| Avg followers per user | 2 |

### 3.4. Model fusion

The final module of our method IMMENSE combines the outputs of the three previous modules, to assign a definitive label to each user. Specifically, in this phase, we train a Multi-Layer Perceptron (MLP) that acts as a meta-model, considering the following input features: the risky and safe autoencoders reconstruction errors $R_R$, $R_S$, along with the label $L^{(con)}$ from the semantic content analysis; the predicted label probabilities of the user being risky or safe, respectively $P_s^{rel}(u)$ and $P_r^{rel}(u)$, from the analysis of social relationships; the predicted probabilities $P_s^{sp}(u)$ and $P_r^{sp}(u)$, of the user being risky or safe, based on the analysis of the user spatial closeness.

The MLP has one hidden layer with the sigmoid activation function, which allows it to capture possible nonlinear dependencies between input and output variables. Since we are dealing with a classification task, the final layer adopts the softmax activation function, which outputs the predicted probabilities of each user to belong to one class or the other. Given that the outputs are probabilistic, we train the MLP using the negative log likelihood as loss function.

We remind that our task is strongly imbalanced. Therefore, also the MLP adopted for the model fusion is trained using one of the strategies outlined in Section 3.2.

## 4. Experiments

In order to evaluate the effectiveness of the proposed method IMMENSE, we performed an extensive set of experiments. In the following subsections, we first provide some details about the considered dataset, then we describe the considered competitor systems and the experimental setting. Finally, we report and discuss the obtained results.

### 4.1. The considered dataset

The dataset for our experimental evaluation was built on the basis of a list of keywords related to radicalism and on a set of known radical/risky posts $D_R$, provided in the context of the Horizon 2020 project CounteR[2] by various Law Enforcement Agencies (LEAs). Using the Twitter API, we retrieved up to 1500 tweets for each keyword, covering the period from February 6th, 2022 to February 6th, 2023. For each author of the downloaded tweets, we then retrieved a list of their followers, limited to a maximum of 1000 due to API usage restrictions. This process allowed us to create a network of social relationships. We finally kept only users who follow more than 5 other users in the obtained list of users, and retrieved up to 20 of the most recent tweets for each user in the network.

To define the ground truth, we adopted the following procedure. We employed a pre-trained Word2Vec model[3] to process the dataset $D_R$, extracting $|D_R|$ risky embeddings. We then aggregated these embeddings to obtain a global risky vector $v_R$, by summation. In the same way, we obtained an embedding for each user, based on his/her posts. We then measured the similarity between $v_R$ and the embedding of each user using cosine similarity. If this similarity exceeded a specified threshold $\delta$, the user was labeled as *risky* in the ground truth, otherwise it was labeled as *safe*. In this way, we quantify how much the user's posts are close to actual risky posts. More formally, for each user $u$, we assigned a label as:

$$label(u) = \begin{cases} risky & \text{if } cosine_similarity(v_u, v_R) \geq \delta \\ safe & \text{otherwise} \end{cases} \tag{11}$$

where $v_u$ is the embedding of the content posted by the user $u$. After analyzing the distribution of similarity values, we set the threshold $\delta$ to 0.88, which results in approximately 7% of users being labeled as *risky*, which can be considered a reasonable proportion in this context according to the LEAs (consider that the corpus is retrieved starting from an expert-defined set of keywords). Finally, we incorporated an additional step that considers the social relationships of users alongside their content. After the initial labeling based on the similarity with the $v_R$ semantic vector, we examined the network of relationships of users labeled as *safe*. Any of these users having more than 10% of their relationships with *risky* users were relabeled as *risky*.

[2] https://counter-project.eu/.
[3] https://code.google.com/archive/p/word2vec/.

**Table 2**
Results obtained by IMMENSE (with class weighting). The best result in terms of F1 is shown in bold.

| IMMENSE: $k_c = k_r = k_s = 128$ | | | | | | | | | | | | |
|---|---|---|---|---|---|---|---|---|---|---|---|---|
| Configuration | | | All users | | | | Safe | | | Risky | | |
| C | R | S | Prec | Rec | F1 | Acc | Prec | Rec | F1 | Prec | Rec | F1 |
| ✓ | | | 0.765 | 0.960 | 0.829 | 0.940 | 1.000 | 0.940 | 0.970 | 0.530 | 0.980 | 0.688 |
| ✓[a] | | | 0.775 | 0.950 | 0.834 | 0.940 | 1.000 | 0.940 | 0.969 | 0.550 | 0.960 | 0.699 |
| | ✓ | | 0.720 | 0.940 | 0.779 | 0.910 | 1.000 | 0.910 | 0.953 | 0.440 | 0.970 | 0.605 |
| | ✓[a] | | 0.700 | 0.935 | 0.750 | 0.900 | 1.000 | 0.890 | 0.940 | 0.400 | 0.980 | 0.560 |
| | | ✓ | 0.775 | 0.955 | 0.836 | 0.940 | 1.000 | 0.940 | 0.969 | 0.550 | 0.970 | 0.702 |
| | | ✓[a] | 0.795 | 0.940 | 0.850 | 0.750 | 0.990 | 0.950 | 0.970 | 0.600 | 0.930 | 0.730 |
| ✓ | ✓ | | 0.875 | 0.975 | 0.918 | **0.980** | 1.000 | 0.980 | **0.990** | 0.750 | 0.970 | 0.846 |
| ✓ | | ✓ | 0.840 | 0.975 | 0.891 | 0.970 | 1.000 | 0.970 | 0.980 | 0.680 | 0.980 | 0.803 |
| | ✓ | ✓ | 0.715 | 0.940 | 0.773 | 0.910 | 1.000 | 0.900 | 0.947 | 0.430 | 0.980 | 0.598 |
| ✓ | ✓ | ✓ | 0.875 | 0.975 | 0.918 | **0.980** | 1.000 | 0.980 | **0.990** | 0.750 | 0.970 | 0.846 |
| IMMENSE: $k_c = k_r = k_s = 256$ | | | | | | | | | | | | |
| Configuration | | | All users | | | | Safe | | | Risky | | |
| C | R | S | Prec | Rec | F1 | Acc | Prec | Rec | F1 | Prec | Rec | F1 |
| ✓ | | | 0.820 | 0.975 | 0.879 | 0.960 | 1.000 | 0.960 | 0.980 | 0.640 | 0.990 | 0.777 |
| ✓[a] | | | 0.940 | 0.900 | 0.919 | **0.980** | 0.990 | 0.990 | **0.990** | 0.890 | 0.810 | 0.848 |
| | ✓ | | 0.745 | 0.940 | 0.802 | 0.930 | 1.000 | 0.930 | 0.964 | 0.490 | 0.950 | 0.640 |
| | ✓[a] | | 0.745 | 0.940 | 0.805 | 0.930 | 1.000 | 0.930 | 0.964 | 0.490 | 0.950 | 0.647 |
| | | ✓ | 0.730 | 0.945 | 0.791 | 0.920 | 1.000 | 0.910 | 0.953 | 0.460 | 0.980 | 0.630 |
| | | ✓[a] | 0.735 | 0.950 | 0.797 | 0.920 | 1.000 | 0.920 | 0.958 | 0.470 | 0.980 | 0.635 |
| ✓ | ✓ | | 0.875 | 0.975 | 0.918 | **0.980** | 1.000 | 0.980 | **0.990** | 0.750 | 0.970 | 0.846 |
| ✓ | | ✓ | 0.850 | 0.980 | 0.902 | 0.970 | 1.000 | 0.970 | 0.985 | 0.700 | 0.990 | 0.820 |
| | ✓ | ✓ | 0.745 | 0.950 | 0.805 | 0.930 | 1.000 | 0.930 | 0.960 | 0.490 | 0.970 | 0.650 |
| ✓ | ✓ | ✓ | 0.880 | 0.975 | 0.921 | **0.980** | 1.000 | 0.980 | **0.990** | 0.760 | 0.970 | 0.852 |
| IMMENSE: $k_c = k_r = k_s = 512$ | | | | | | | | | | | | |
| Configuration | | | All users | | | | Safe | | | Risky | | |
| C | R | S | Prec | Rec | F1 | Acc | Prec | Rec | F1 | Prec | Rec | F1 |
| ✓ | | | 0.850 | 0.975 | 0.901 | 0.970 | 1.000 | 0.970 | 0.985 | 0.700 | 0.980 | 0.817 |
| ✓[a] | | | 0.850 | 0.975 | 0.901 | 0.970 | 1.000 | 0.970 | 0.985 | 0.700 | 0.980 | 0.817 |
| | ✓ | | 0.710 | 0.940 | 0.769 | 0.910 | 1.000 | 0.900 | 0.950 | 0.420 | 0.980 | 0.588 |
| | ✓[a] | | 0.705 | 0.935 | 0.760 | 0.900 | 1.000 | 0.890 | 0.942 | 0.410 | 0.980 | 0.578 |
| | | ✓ | 0.710 | 0.940 | 0.769 | 0.900 | 1.000 | 0.900 | 0.947 | 0.420 | 0.980 | 0.590 |
| | | ✓[a] | 0.690 | 0.935 | 0.745 | 0.890 | 1.000 | 0.880 | 0.940 | 0.380 | 0.990 | 0.549 |
| ✓ | ✓ | | 0.880 | 0.975 | 0.921 | **0.980** | 1.000 | 0.980 | **0.990** | 0.760 | 0.970 | 0.852 |
| ✓ | | ✓ | 0.880 | 0.975 | 0.918 | **0.980** | 1.000 | 0.980 | **0.990** | 0.750 | 0.970 | 0.846 |
| | ✓ | ✓ | 0.725 | 0.945 | 0.785 | 0.910 | 1.000 | 0.910 | 0.953 | 0.450 | 0.980 | 0.617 |
| ✓ | ✓ | ✓ | 0.890 | 0.980 | **0.930** | **0.980** | 1.000 | 0.980 | **0.990** | 0.780 | 0.980 | **0.870** |

[a] For IMMENSE configurations represents the adoption of its C, R, and S modules without the fusion module.

In Table 1, we show some quantitative information about the dataset. From the table, we can observe that the information about the geographical position is available only for a small subset of users.

### 4.2. Experimental setting and competitors

Our experiments were carried out by splitting the dataset into training (80% of the users) and testing (20% of the users) sets. The split was made so that the two sets have no common user. In this way, we properly evaluate the inductive capabilities of the system, namely, its ability to effectively generalize to new users, who were not observed during the training phase.

We experimented with three different values of the embedding dimensionality for the textual content $k_c$, for the network of relationships $k_r$, and for the spatial dimension $k_s$. Specifically, we evaluated the following settings: $k_c = k_r = k_s = 128$, $k_c = k_r = k_s = 256$, and $k_c = k_r = k_s = 512$. We choose these values for embedding dimensions based on their widespread use in related work [23,48–51]. We did not consider configurations with different sizes for different perspectives in order to always provide the same *a-priori importance* to all perspectives and leave the fusion module to properly assess and combine their contribution. We report the results achieved using both the proposed strategies to handle class imbalance (see Section 3.2), namely, *class weighting* and *focal loss*. For the latter, we used the values suggested in the original paper [47] for its hyperparameters, i.e., $\gamma = 2.0$ and $\alpha = 0.25$.

To specifically evaluate the contribution provided by each perspective (content - **C**, relationships - **R**, and spatial information - **S**), we also measured the performance considering different combinations thereof.

**Table 3**
Results obtained by IMMENSE (with focal loss). The best result in terms of F1 is shown in bold.

| IMMENSE: $k_c = k_r = k_s = 128$ | | | | | | | | | | | | |
|---|---|---|---|---|---|---|---|---|---|---|---|---|
| Configuration | | | All users | | | | Safe | | | Risky | | |
| C | R | S | Prec | Rec | F1 | Acc | Prec | Rec | F1 | Prec | Rec | F1 |
| ✓ | | | 0.965 | 0.925 | 0.944 | **0.990** | 0.990 | 1.000 | **0.995** | 0.940 | 0.850 | 0.893 |
| ✓[a] | | | 0.775 | 0.950 | 0.834 | 0.940 | 1.000 | 0.940 | 0.969 | 0.550 | 0.960 | 0.699 |
| | ✓ | | 0.835 | 0.930 | 0.875 | 0.960 | 0.990 | 0.970 | 0.980 | 0.680 | 0.890 | 0.771 |
| | ✓[a] | | 0.820 | 0.945 | 0.870 | 0.960 | 0.990 | 0.960 | 0.975 | 0.650 | 0.930 | 0.765 |
| | | ✓ | 0.895 | 0.920 | 0.907 | 0.980 | 0.990 | 0.980 | 0.985 | 0.800 | 0.860 | 0.829 |
| | | ✓[a] | 0.895 | 0.920 | 0.907 | 0.980 | 0.990 | 0.980 | 0.985 | 0.800 | 0.860 | 0.829 |
| ✓ | ✓ | | 0.965 | 0.950 | 0.955 | **0.990** | 0.990 | 1.000 | 0.990 | 0.940 | 0.900 | 0.920 |
| ✓ | | ✓ | 0.960 | 0.940 | 0.947 | **0.990** | 0.990 | 1.000 | 0.990 | 0.930 | 0.880 | 0.904 |
| | ✓ | ✓ | 0.910 | 0.930 | 0.920 | 0.980 | 0.990 | 0.990 | 0.990 | 0.830 | 0.870 | 0.850 |
| ✓ | ✓ | ✓ | 0.980 | 0.850 | 0.900 | 0.980 | 0.980 | 1.000 | 0.990 | 0.980 | 0.700 | 0.810 |
| IMMENSE: $k_c = k_r = k_s = 256$ | | | | | | | | | | | | |
| Configuration | | | All users | | | | Safe | | | Risky | | |
| C | R | S | Prec | Rec | F1 | Acc | Prec | Rec | F1 | Prec | Rec | F1 |
| ✓ | | | 0.965 | 0.940 | 0.952 | **0.990** | 0.990 | 1.000 | **0.995** | 0.940 | 0.880 | 0.909 |
| ✓[a] | | | 0.940 | 0.900 | 0.919 | 0.980 | 0.990 | 0.990 | 0.990 | 0.890 | 0.810 | 0.848 |
| | ✓ | | 0.820 | 0.905 | 0.856 | 0.960 | 0.990 | 0.970 | 0.980 | 0.650 | 0.840 | 0.733 |
| | ✓[a] | | 0.815 | 0.910 | 0.855 | 0.960 | 0.990 | 0.970 | 0.980 | 0.640 | 0.850 | 0.730 |
| | | ✓ | 0.870 | 0.895 | 0.882 | 0.970 | 0.990 | 0.980 | 0.985 | 0.750 | 0.810 | 0.779 |
| | | ✓[a] | 0.870 | 0.880 | 0.875 | 0.970 | 0.980 | 0.980 | 0.980 | 0.760 | 0.780 | 0.770 |
| ✓ | ✓ | | 0.960 | 0.955 | 0.957 | **0.990** | 0.990 | 1.000 | **0.995** | 0.930 | 0.910 | 0.920 |
| ✓ | | ✓ | 0.875 | 0.965 | 0.914 | 0.970 | 1.000 | 0.980 | 0.990 | 0.750 | 0.950 | 0.838 |
| | ✓ | ✓ | 0.865 | 0.915 | 0.888 | 0.970 | 0.990 | 0.980 | 0.985 | 0.740 | 0.850 | 0.791 |
| ✓ | ✓ | ✓ | 0.965 | 0.955 | 0.960 | **0.990** | 0.990 | 1.000 | **0.995** | 0.940 | 0.910 | 0.925 |
| IMMENSE: $k_c = k_r = k_s = 512$ | | | | | | | | | | | | |
| Configuration | | | All users | | | | Safe | | | Risky | | |
| C | R | S | Prec | Rec | F1 | Acc | Prec | Rec | F1 | Prec | Rec | F1 |
| ✓ | | | 0.970 | 0.945 | 0.957 | **0.990** | 0.990 | 1.000 | **0.995** | 0.950 | 0.890 | 0.919 |
| ✓[a] | | | 0.850 | 0.975 | 0.901 | 0.970 | 1.000 | 0.970 | 0.985 | 0.700 | 0.980 | 0.817 |
| | ✓ | | 0.780 | 0.935 | 0.837 | 0.950 | 0.990 | 0.950 | 0.970 | 0.570 | 0.920 | 0.704 |
| | ✓[a] | | 0.755 | 0.940 | 0.812 | 0.930 | 1.000 | 0.930 | 0.964 | 0.510 | 0.950 | 0.660 |
| | | ✓ | 0.845 | 0.905 | 0.870 | 0.960 | 0.990 | 0.970 | 0.980 | 0.700 | 0.840 | 0.760 |
| | | ✓[a] | 0.810 | 0.930 | 0.857 | 0.960 | 0.990 | 0.960 | 0.975 | 0.630 | 0.900 | 0.740 |
| ✓ | ✓ | | 0.905 | 0.960 | 0.930 | 0.980 | 1.000 | 0.980 | 0.990 | 0.810 | 0.940 | 0.870 |
| ✓ | | ✓ | 0.970 | 0.955 | 0.962 | **0.990** | 0.990 | 1.000 | **0.995** | 0.950 | 0.910 | 0.930 |
| | ✓ | ✓ | 0.835 | 0.925 | 0.875 | 0.960 | 0.990 | 0.970 | 0.980 | 0.680 | 0.880 | 0.770 |
| ✓ | ✓ | ✓ | 0.970 | 0.965 | **0.967** | **0.990** | 0.990 | 1.000 | **0.995** | 0.950 | 0.930 | **0.940** |

[a] For IMMENSE configurations represents the adoption of its C, R, and S modules without the fusion module.

We also performed comparative experiments considering each single module implemented in IMMENSE.

The results obtained by IMMENSE were compared with those achieved by the following state-of-the-art systems, namely:

- *Mr-SBC* [39], that represents the dataset as database tables and adopts a probabilistic approach for the node classification task.
- *Re3py* [38], an approach based on ensembles of relational decision trees. Its peculiarity is the ability to construct new features dynamically by navigating foreign key paths and employing an iterative feature aggregation strategy.
- *Heterogeneous Graph Transformer (HGT)* [24], where, as suggested by the authors, node features are extracted from the content using the pre-trained XLNet language model [52], which identifies embeddings of dimension 768.
- *Rio-GNN* [26], adopting the same procedure followed by the authors: Rio-GNN is used for computing node embeddings and the classification is done by an MLP. We set the features initially associated with nodes to the same features used for IMMENSE, and its parameter *n_emb* (dimension of the node embeddings to learn) to the same value as the initial feature vector.
- *SAIRUS* [1], a transductive framework for user classification in social networks, able to consider the posted textual content, user relationships, and users' spatial closeness.

It is worth noting that both *Mr-SBC* and *Re3py* represent user content through a bag-of-words approach, but represent relationships differently: *Mr-SBC* uses foreign key constraints, while *Re3py* relies on first-order predicates. Unlike our method, these two approaches do not exploit embeddings to capture the semantics of the content. HGT and Rio-GNN, on the other hand, adopt an embedding-based approach for representing the nodes in the network. For both HGT and Rio-GNN, we applied the same weighting schema used by IMMENSE (see Eq. (3)) to their respective loss functions, to account for class imbalance.

It is important to note that SAIRUS, being transductive, requires access to the complete network of relationships and geographical information during training, and cannot provide predictions for new users, who have not been observed in those networks, without an additional training phase on the network of relationships and on that representing the geographical information. IMMENSE, on the other hand, enables generalization to completely new networks in the prediction phase. This aspect makes the comparison inherently unfair in favor of SAIRUS, since it is aware of the users in the testing set during the training, while IMMENSE is purposely made unaware of them. However, such a comparison allows us to assess the performance of IMMENSE in such a more challenging scenario, compared with its closest transductive competitor.

All the experiments were performed on a workstation equipped with a NVIDIA GeForce Titan X GPU, an Intel Xeon E5-1650-v3 CPU, and 64 GB of RAM. As evaluation measures, we consider precision, recall, accuracy, and F1-score, computed both on the entire set of users and for each class separately.

**Table 4**
Results obtained by inductive competitors. The best result in terms of F1 is shown in bold.

| MrSBC | | | | | | | | | | | | |
|---|---|---|---|---|---|---|---|---|---|---|---|---|
| Configuration | | | All users | | | | Safe | | | Risky | | |
| C | R | S | Prec | Rec | F1 | Acc | Prec | Rec | F1 | Prec | Rec | F1 |
| ✓ | | | 0.465 | 0.500 | 0.482 | 0.930 | 0.930 | 1.000 | **0.964** | 0.000 | 0.000 | 0.000 |
| ✓ | ✓ | | 0.465 | 0.500 | 0.482 | 0.930 | 0.930 | 1.000 | **0.964** | 0.000 | 0.000 | 0.000 |
| ✓ | | ✓ | 0.465 | 0.500 | 0.482 | 0.930 | 0.930 | 1.000 | **0.964** | 0.000 | 0.000 | 0.000 |
| ✓ | ✓ | ✓ | 0.465 | 0.500 | 0.482 | 0.930 | 0.930 | 1.000 | **0.964** | 0.000 | 0.000 | 0.000 |
| Re3py | | | | | | | | | | | | |
| Configuration | | | All users | | | | Safe | | | Risky | | |
| C | R | S | Prec | Rec | F1 | Acc | Prec | Rec | F1 | Prec | Rec | F1 |
| ✓ | | | 0.465 | 0.500 | 0.482 | 0.930 | 0.930 | 1.000 | **0.964** | 0.000 | 0.000 | 0.000 |
| ✓ | ✓ | | 0.465 | 0.500 | 0.482 | 0.930 | 0.930 | 1.000 | **0.964** | 0.000 | 0.000 | 0.000 |
| ✓ | | ✓ | 0.465 | 0.500 | 0.482 | 0.930 | 0.930 | 1.000 | **0.964** | 0.000 | 0.000 | 0.000 |
| ✓ | ✓ | ✓ | 0.465 | 0.500 | 0.482 | 0.930 | 0.930 | 1.000 | **0.964** | 0.000 | 0.000 | 0.000 |
| HGT | | | | | | | | | | | | |
| Configuration | | | All users | | | | Safe | | | Risky | | |
| | | | Prec | Rec | F1 | Acc | Prec | Rec | F1 | Prec | Rec | F1 |
| – | | | 0.680 | 0.580 | 0.610 | 0.930 | 0.940 | 0.980 | 0.960 | 0.420 | 0.180 | 0.252 |
| RIO-GNN | | | | | | | | | | | | |
| Configuration | | | All users | | | | Safe | | | Risky | | |
| | | | Prec | Rec | F1 | Acc | Prec | Rec | F1 | Prec | Rec | F1 |
| $n_emb = 128$ | | | 0.535 | 0.515 | 0.447 | 0.510 | 0.500 | 0.860 | 0.632 | 0.570 | 0.170 | 0.262 |
| $n_emb = 256$ | | | 0.790 | 0.775 | **0.773** | 0.780 | 0.720 | 0.880 | 0.792 | 0.860 | 0.670 | **0.753** |
| $n_emb = 512$ | | | 0.630 | 0.605 | 0.594 | 0.610 | 0.670 | 0.410 | 0.509 | 0.590 | 0.800 | 0.679 |

### 4.3. Results and discussion

In Tables 2 and 3, we show the results obtained by IMMENSE with the class weighting and the focal loss strategies, respectively, while in Table 4 we show the results of all its inductive competitors. The results refer to multiple configurations, considering the content (C), the social relationships (R), the spatial closeness (S), and combinations thereof, when made possible by each considered approach. The symbol[a] for IMMENSE configurations represents the adoption of its C, R, and S modules without the final fusion module. Looking at the results, it is immediately noticeable that competitors fail to achieve satisfactory results across all evaluated configurations. Specifically, *MrSBC* and *Re3py* exhibit a strong bias towards the *safe* class, classifying all users as *safe* in all the considered configurations (see the recall equal to 0 for the *risky* class). Both competitors clearly suffer from data unbalancing issues, being not able to provide a good trade-off in the predictive performance over both classes. A further motivation behind poor performances of these competitors may be their bag-of-words representation for the content, that can be considered suboptimal as it fails to capture the semantics.

On the other hand, *HGT* manages to correctly identify some *risky* users (recall on the risky class equal to 0.18), but it still tends to classify most of users as *safe*. Therefore, its performance still appears affected by class imbalance, despite the adoption of a weighted loss function during the training.

*Rio-GNN* provides more useful predictions. The best results are achieved with $n_emb = 256$ (average F1 equal to 0.773), but they still remain below those obtained by our method IMMENSE. We believe that such performances exhibited by *Rio-GNN* depend on the fact that the content is taken into account exclusively through the initial features associated with the nodes in the network, without a dedicated module for the analysis of the semantics of the content. On the contrary, IMMENSE has a dedicated component for each perspective (i.e., content, relations and spatial closeness).

Focusing on IMMENSE (see Tables 2 and 3), it is clear that the obtained results are far better than those achieved by all the competitors. In general, IMMENSE achieves the best results when all the three perspectives are considered, although in a few cases, some configurations considering two perspectives yield the same macro results. However, the results obtained for the *risky* class (that is the most interesting class for our task) reveal that taking all perspectives into account leads to the best performance with all the configurations, when class weighting is adopted, and with an embedding dimensionality of 256 or 512, when focal loss is adopted. We can also notice that considering one single perspective, either directly (see the configuration with the[a] or going through the final fusion step, generally provides suboptimal performances. Moreover, considering the content always appears to be fundamental for accurate predictions, even if clearly complemented by the perspectives modeling social relationships and spatial proximity among users. Notably, the F1 performances increase in 5 out of 6 cases when complementing the information provided by the posted content with that of the spatial dimension. This is particularly interesting, considering that the spatial information is available for a very limited number of users (see Table 1). This result also outlines the ability of IMMENSE of properly managing possible sparsity issues in the available data.

It can also be noticed that IMMENSE provides balanced performances over the *risky* and *safe* classes. These results highlight the capability of IMMENSE of properly exploiting multiple, complementary, perspectives as well as handling class imbalance, which is typical in the considered context.

As expected, increasing the embedding dimensionality tends to improve performance. However, the results obtained with embedding sizes of 128 and 256 are still highly competitive, significantly outperforming all baselines across all configurations.

In Table 5, we separately show the results obtained by SAIRUS, which can be considered the closest competitor to IMMENSE. We remind that such results are achieved in a transductive setting, and are, in principle, not comparable with those achieved by IMMENSE in the inductive setting, since SAIRUS is made aware of users for which it will need to provide a label during the prediction phase. Despite this inherent advantage provided to SAIRUS, in Table 6 we can see that IMMENSE outperforms SAIRUS in almost all the cases. In particular, it reaches an improvement of 9.41% in terms of F1 on all users and of 20.83% on *risky* users, with a dimensionality of 512, when class weighting is adopted, and an improvement of 14.12% in terms of F1 on all users and of 30.54% on *risky* users, with a dimensionality of 512, when focal loss is adopted. This gain comes with no additional computational costs for IMMENSE. Actually, it exhibits up to 76.93% improvement in terms of running times during the prediction phase (see the last column of Table 6) since it does not need to re-train the node embedding models on the graph used for the inference phase. Table 7 provides additional details on both training and inference running times. As it can be noticed, IMMENSE is more efficient then SAIRUS for both phases, and exhibits very low running times for the inference phase (at most 12.4 s for a testing network of ∼7500 users), making it practically adoptable in real-world scenarios.

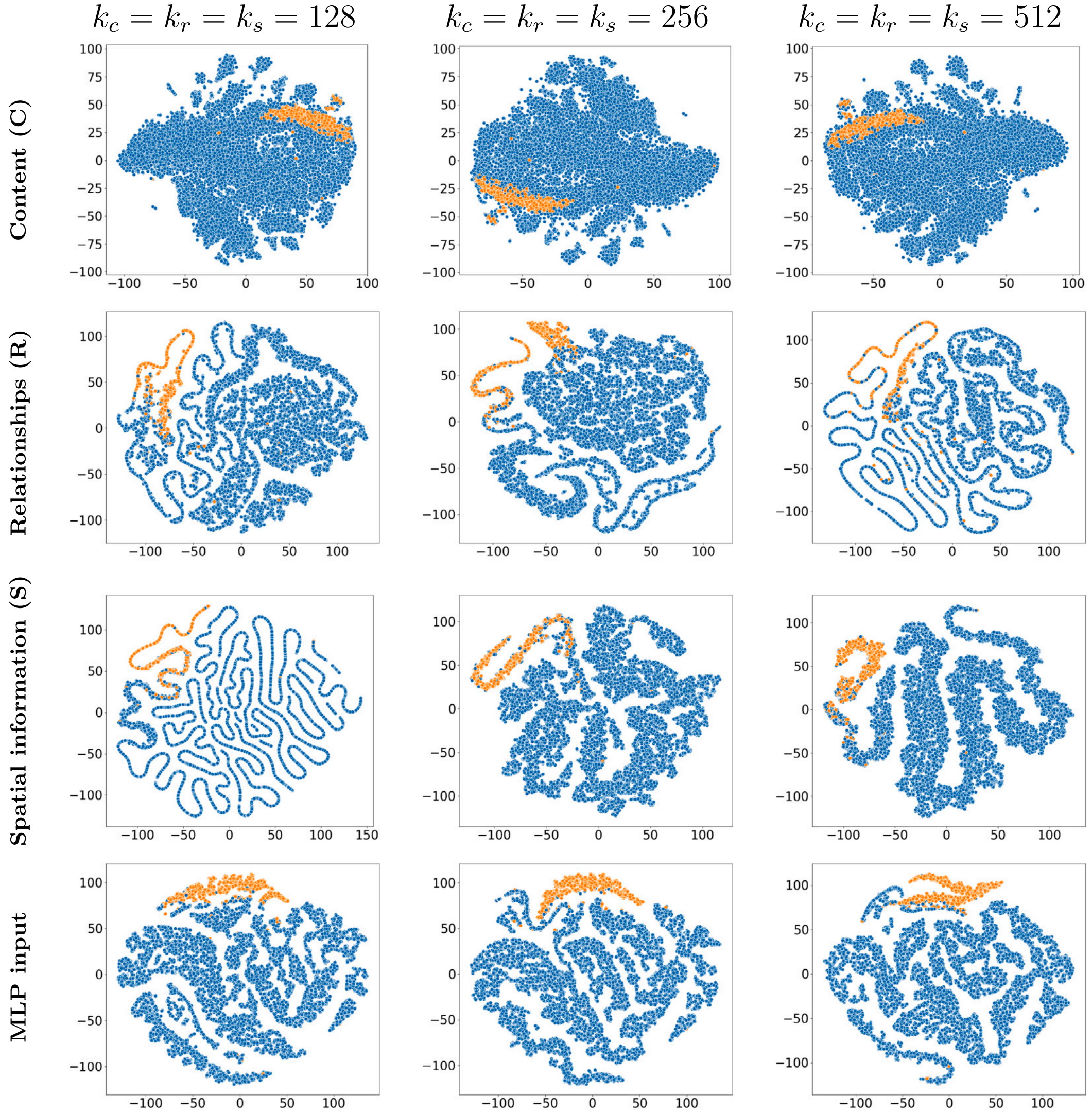

**Fig. 3.** Graphical representation through t-SNE of the IMMENSE (with focal loss) embeddings for each perspective and for the input of the fusion module (MLP). Orange data points correspond to *risky* users, while blue data points correspond to *safe* users. (For interpretation of the references to color in this figure legend, the reader is referred to the web version of this article.)

**Table 5**
Results obtained by the transductive competitor SAIRUS.

| SAIRUS | | | | | | | | | | |
|---|---|---|---|---|---|---|---|---|---|---|
| Configuration | All users | | | | Safe | | | Risky | | |
| | Prec | Rec | F1 | Acc | Prec | Rec | F1 | Prec | Rec | F1 |
| $k_c = k_r = k_s = 128$ | 0.850 | 0.960 | 0.900 | 0.970 | 1.000 | 0.970 | 0.980 | 0.710 | 0.940 | 0.810 |
| $k_c = k_r = k_s = 256$ | 0.870 | 0.910 | 0.890 | 0.970 | 0.990 | 0.980 | 0.980 | 0.750 | 0.850 | 0.800 |
| $k_c = k_r = k_s = 512$ | 0.970 | 0.780 | 0.850 | 0.970 | 0.970 | 1.000 | 0.980 | 0.970 | 0.570 | 0.720 |

All these results prove the capability of IMMENSE in providing accurate predictions for this specific task in social networks, properly exploiting the complementary information conveyed by three different perspectives. The specific comparison with SAIRUS also emphasized its ability not only to outperform it from a mere viewpoint of the accuracy of the predictions, but also in terms of sustainability, due to the significant reduction of the inference time achieved through the proposed inductive approach.

We finally performed a specific analysis to assess the contribution of each perspective in discriminating between *risky* and *safe* users. Specifically, in Fig. 3 we report several t-SNE plots [53] depicting the 2D projection of the embeddings identified by the modules devoted to the analysis of the posted content (C), of social relationships (R), and of spatial information (S). We also report the 2D projection of the 7-dimensional feature space that represents the input of the final fusion module based on the MLP. For the sake of compactness, we only report the plots obtained when the focal loss is adopted. From the Figure, we can observe that each perspective alone already provides some useful information for separating *risky* and *safe* users. However, the MLP input exhibits a very neat separation between the two classes, with only 7 features. These visual results confirm that the three perspectives, subsequently properly combined by the MLP, provide complementary information to the fusion module to better discriminate between *risky* and *safe* users, thus confirming our initial intuition behind IMMENSE.

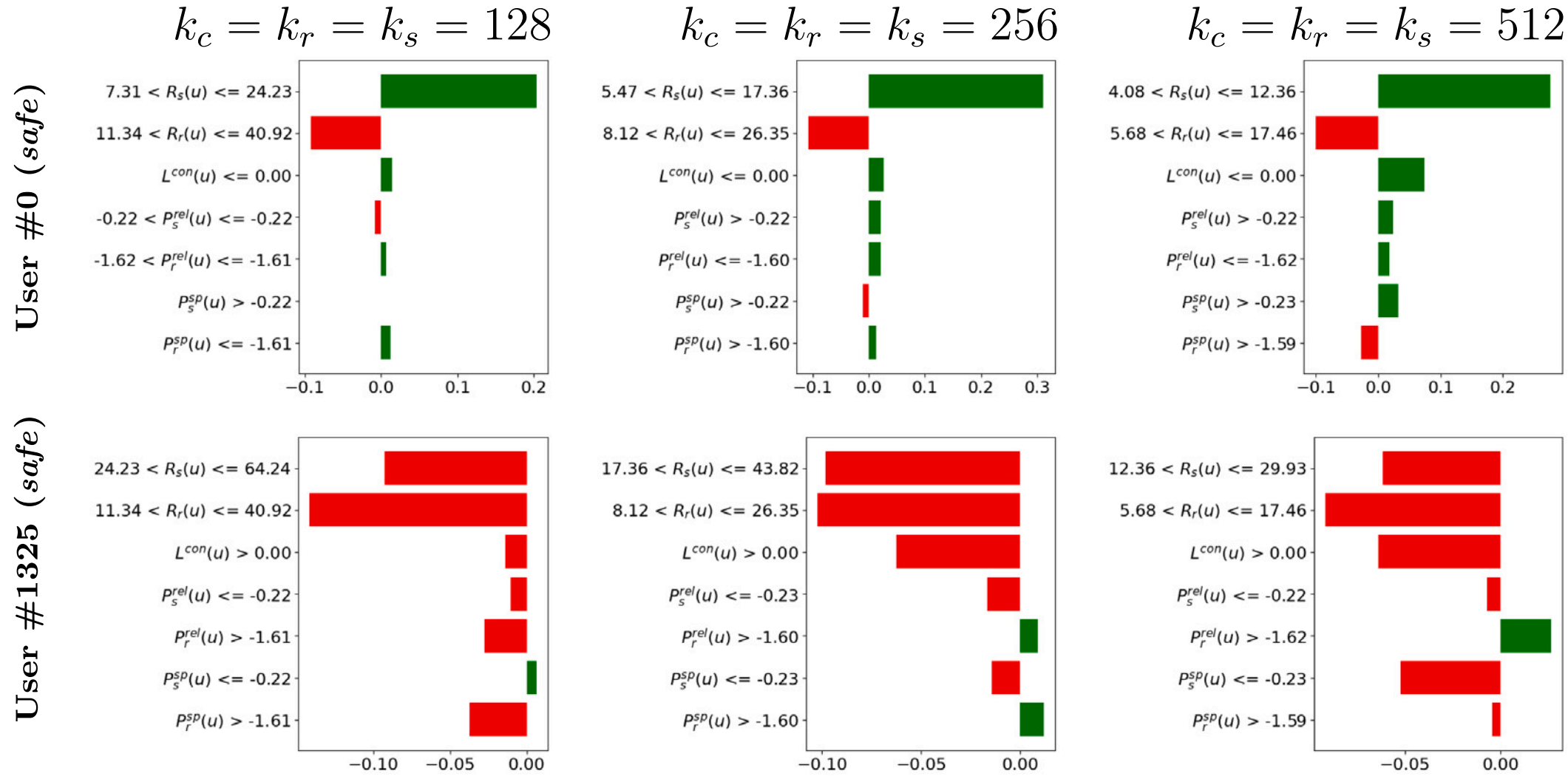

**Fig. 4.** Graphical representation through LIME of the importance of each input feature of the fusion module (MLP) of IMMENSE (with focal loss), for 2 selected *safe* users. User #0 is correctly classified by IMMENSE as *safe*, while user #1325 is wrongly classified by IMMENSE as *risky*. (For interpretation of the references to color in this figure legend, the reader is referred to the web version of this article.)

**Table 6**
Improvements of IMMENSE over SAIRUS. The improvement in terms of running times (running time reduction) refers to the inference phase.

| IMMENSE (with class weighting) vs. SAIRUS | | | | | | | | | | | | |
|---|---|---|---|---|---|---|---|---|---|---|---|---|
| Configuration | All users | | | | Safe | | | Risky | | | | |
| | Prec | Rec | F1 | Acc | Prec | Rec | F1 | Prec | Rec | F1 | Time | |
| $k_c = k_r = k_s = 128$ | 2.94% | 1.56% | 1.99% | 1.03% | 0.00% | 1.03% | 1.01% | 5.63% | 3.19% | 4.44% | 75.66% | |
| $k_c = k_r = k_s = 256$ | 1.15% | 7.14% | 3.49% | 1.03% | 1.01% | 0.00% | 1.01% | 1.33% | 14.12% | 6.53% | 73.43% | |
| $k_c = k_r = k_s = 512$ | −8.25% | 25.64% | 9.41% | 1.03% | 3.09% | −2.00% | 1.01% | −19.59% | 71.93% | 20.83% | 75.93% | |
| **IMMENSE (with focal loss) vs. SAIRUS** | | | | | | | | | | | | |
| Configuration | All users | | | | Safe | | | Risky | | | | |
| | Prec | Rec | F1 | Acc | Prec | Rec | F1 | Prec | Rec | F1 | Time | |
| $k_c = k_r = k_s = 128$ | 15.29% | −11.46% | 0.00% | 1.03% | −2.00% | 3.09% | 1.02% | 38.03% | −25.53% | 0.00% | 71.36% | |
| $k_c = k_r = k_s = 256$ | 11.49% | 4.40% | 7.87% | 2.06% | 0.00% | 2.04% | 1.53% | 25.33% | 7.06% | 15.59% | 76.70% | |
| $k_c = k_r = k_s = 512$ | 0.00% | 23.08% | 14.12% | 2.06% | 2.06% | 0.00% | 1.53% | −2.06% | 63.16% | 30.54% | 76.93% | |

**Table 7**
Training and inference times of IMMENSE and SAIRUS for the whole set of users in the training and testing sets, respectively. Note that SAIRUS requires to partially re-train the model during the inference because of its transductive nature.

| Configuration | Training time (s) | | | Inference time (s) | | |
|---|---|---|---|---|---|---|
| | SAIRUS | IMMENSE (class weighting) | IMMENSE (focal loss) | SAIRUS | IMMENSE (class weighting) | IMMENSE (focal loss) |
| $k_c = k_r = k_s = 128$ | 486.3 | 357.2 | 416.3 | 46.4 | 10.4 | 11.2 |
| $k_c = k_r = k_s = 256$ | 612.5 | 395.8 | 454.8 | 46.4 | 11.2 | 12.4 |
| $k_c = k_r = k_s = 512$ | 874.0 | 536.4 | 597.3 | 51.3 | 11.3 | 12.4 |

Furthermore, we also selected two safe users and two risky users, and plotted their corresponding LIME importance scores [54] in Figs. 4 and 5. In particular, Fig. 4 shows the LIME plots for one safe user (User #0) correctly classified by IMMENSE and one of the few *safe* users (User #1325) wrongly classified by IMMENSE as *risky*. As we can see, for User #0, especially with embedding size 512, the consistency (with the true class) of the meta-features (input features of the MLP) denoted by the green bar counterbalances the mistakes of other meta-features, with the reconstruction error of the safe autoencoder being the most contributing feature for a correct classification. For User #1325, we see that the error in the classification is mainly due to all perspectives. A clear imperfect classification based on the content is not counterbalanced by the other perspectives, which, instead, reinforce the error. We note that this example was selected on purpose, and this behavior is not commonly observed in other examples, as F1 and accuracy scores confirm.

Fig. 5 highlights additional noteworthy cases involving *risky* users. Notably, for User #24, the correct classification into the *risky* class is strongly supported by almost all meta-features. Only two of them show a very weak inclination towards classifying the user as *safe*. Instead, for User #689 a strong mistake due to the reconstruction error of the *safe* autoencoder, which appears lower than that of the *risky* autoencoder, is not counterbalanced by other meta-features, leading to a wrong classification. Again, this example was deliberately selected to illustrate such behavior, which is not commonly observed across the dataset.

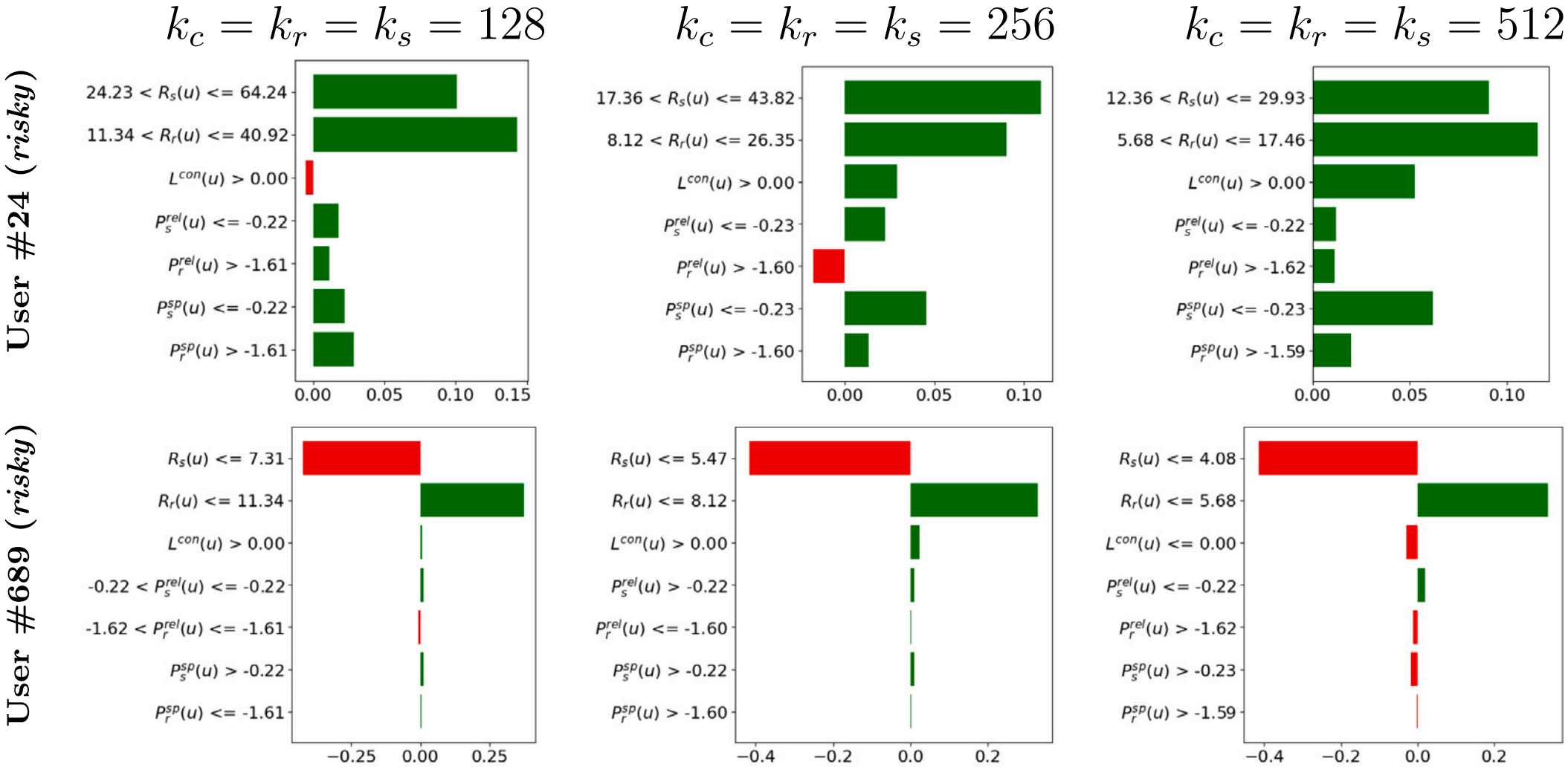

**Fig. 5.** Graphical representation through LIME of the importance of each input feature of the fusion module (MLP) of IMMENSE (with focal loss), for 2 selected *risky* users. User #24 is correctly classified by IMMENSE as *risky*, while user #689 is wrongly classified by IMMENSE as *safe*.

## 5. Conclusions

In this paper we presented IMMENSE, an inductive learning method for the identification of risky users in social networks. IMMENSE can effectively classify unseen nodes by leveraging three perspectives: the semantics of the content posted by users, their social relationships, and their spatial closeness.

Our evaluation on a real-world dataset demonstrated that IMMENSE is able to outperform four state-of-the-art inductive competitors, which struggle with network sparsity and data imbalance. IMMENSE also proved to outperform its closest competitor SAIRUS, even if the latter was run in a more advantageous setting (i.e., transductive), where nodes in the testing set are known in advance during the training phase. Such improvements were clear both in terms of predictive accuracy and in terms of prediction time. These results suggest that IMMENSE can effectively be used in real-world environments by Law Enforcement Agencies to counter negative phenomena in social networks.

For future work, we plan to enhance IMMENSE by incorporating temporal analysis, which would allow tracking changes in user behaviors over time. This aspect could support the detection of users with a safe history who suddenly begin posting negative content or joining risky communities.

## CRediT authorship contribution statement

**Francesco Benedetti:** Writing – original draft, Validation, Software, Methodology, Investigation, Formal analysis, Data curation, Conceptualization. **Antonio Pellicani:** Writing – review & editing, Validation, Methodology, Investigation, Formal analysis, Data curation, Conceptualization. **Gianvito Pio:** Writing – review & editing, Validation, Methodology, Formal analysis, Data curation, Conceptualization. **Michelangelo Ceci:** Writing – review & editing, Validation, Supervision, Funding acquisition, Formal analysis, Conceptualization.

## Code availability

The proposed method IMMENSE is open source and publicly available on Github: https://github.com/itsfrank98/immense.

## Declaration of competing interest

The authors have declared no conflict of interest

## Acknowledgments

The authors acknowledge the support of the European Commission through the H2020 Project "CounteR - Privacy-First Situational Awareness Platform for Violent Terrorism and Crime Prediction, Counter Radicalization and Citizen Protection" (Grant N. 101021607).

This work was partially supported by the project FAIR - Future AI Research (PE00000013), spoke 6 - Symbiotic AI, under the NRRP MUR program funded by the NextGenerationEU.

## Data availability

The data used in this paper will be made available upon request.

## References

[1] A. Pellicani, G. Pio, D. Redavid, M. Ceci, SAIRUS: Spatially-aware identification of risky users in social networks, Inf. Fusion 92 (2023) 435–449, http://dx.doi.org/10.1016/J.INFFUS.2022.11.029.

[2] R.A. Igawa, S. Barbon Jr., K.C.S. Paulo, G.S. Kido, R.C. Guido, M.L.P. Júnior, I.N. da Silva, Account classification in online social networks with lbca and wavelets, Inform. Sci. 332 (2016) 72–83.

[3] W. Wu, M. Ghazali, S. Hazlin Huspi, A review of user profiling based on social networks, IEEE Access 12 (2024) 122642–122670, http://dx.doi.org/10.1109/ACCESS.2024.3430987.

[4] K. Balasubramaniam, S. Vidhya, N. Jayapandian, K. Ramya, M. Poongodi, M. Hamdi, G.B. Tunze, et al., Social network user profiling with multilayer semantic modeling using ego network, Int. J. Inf. Technol. Web Eng. (IJITWE) 17 (1) (2022) 1–14.

[5] H. Ko, S. Lee, Y. Park, A. Choi, A survey of recommendation systems: recommendation models, techniques, and application fields, Electronics 11 (1) (2022) 141.

[6] D. Li, H. Liu, Z. Zhang, K. Lin, S. Fang, Z. Li, N.N. Xiong, CARM: Confidence-aware recommender model via review representation learning and historical rating behavior in the online platforms, Neurocomputing 455 (2021) 283–296.

[7] H. Liu, C. Zheng, D. Li, Z. Zhang, K. Lin, X. Shen, N.N. Xiong, J. Wang, Multi-perspective social recommendation method with graph representation learning, Neurocomputing 468 (2022) 469–481.

[8] H. Liu, C. Zheng, D. Li, X. Shen, K. Lin, J. Wang, Z. Zhang, Z. Zhang, N.N. Xiong, EDMF: Efficient deep matrix factorization with review feature learning for industrial recommender system, IEEE Trans. Ind. Inform. 18 (7) (2022) 4361–4371.

[9] U.N.I. Crime, J.R.I. (UNICRI), Stop the virus of disinformation: The malicious use of social media by terrorist, violent extremist and criminal groups during the covid-19 pandemic - november 2020, 2020.

[10] R. Thompson, Radicalization and the use of social media, J. Strat. Secur. 4 (4) (2011) 167–190.

[11] B. Evkoski, A. Pelicon, I. Mozetic, N. Ljubesic, P.K. Novak, Retweet communities reveal the main sources of hate speech, 2021, CoRR abs/2105.14898. arXiv:2105.14898.

[12] H.S. Alatawi, A.M. Alhothali, K.M. Moria, Detecting white supremacist hate speech using domain specific word embedding with deep learning and BERT, IEEE Access 9 (2021) 106363–106374, http://dx.doi.org/10.1109/ACCESS.2021.3100435.

[13] M. Ji, Y. Sun, M. Danilevsky, J. Han, J. Gao, Graph regularized transductive classification on heterogeneous information networks, in: J.L. Balcázar, F. Bonchi, A. Gionis, M. Sebag (Eds.), Machine Learning and Knowledge Discovery in Databases, European Conference, ECML PKDD 2010, Barcelona, Spain, September 20-24, 2010, Proceedings, Part I, in: Lecture Notes in Computer Science, vol. 6321, Springer, 2010, pp. 570–586, http://dx.doi.org/10.1007/978-3-642-15880-3_42.

[14] L. Berkani, S. Belkacem, M. Ouafi, A. Guessoum, Recommendation of users in social networks: A semantic and social based classification approach, Expert. Syst. J. Knowl. Eng. 38 (2) (2021) http://dx.doi.org/10.1111/EXSY.12634.

[15] C. Desrosiers, G. Karypis, Within-network classification using local structure similarity, in: W. Buntine, M. Grobelnik, D. Mladenić, J. Shawe-Taylor (Eds.), Machine Learning and Knowledge Discovery in Databases, Springer Berlin Heidelberg, Berlin, Heidelberg, 2009, pp. 260–275.

[16] O. Chapelle, V. Vapnik, J. Weston, Transductive inference for estimating values of functions, in: S.A. Solla, T.K. Leen, K. Müller (Eds.), Advances in Neural Information Processing Systems 12, [NIPS Conference, Denver, Colorado, USA, November 29 - December 4 1999], The MIT Press, 1999, pp. 421–427.

[17] Z. Xue, Z. Zhang, H. Liu, S. Yang, S. Han, Learning knowledge graph embedding with multi-granularity relational augmentation network, Expert Syst. Appl. 233 (2023) 120953.

[18] Z. Zhang, Z. Li, H. Liu, N.N. Xiong, Multi-scale dynamic convolutional network for knowledge graph embedding, IEEE Trans. Knowl. Data Eng. 34 (5) (2022) 2335–2347.

[19] Z. Li, H. Liu, Z. Zhang, T. Liu, N.N. Xiong, Learning knowledge graph embedding with heterogeneous relation attention networks, IEEE Trans. Neural Netw. Learn. Syst. 33 (8) (2022) 3961–3973.

[20] Z. Xue, Z. Zhang, H. Liu, Z. Li, S. Han, E. Zhang, Mhrn: A multi-perspective hierarchical relation network for knowledge graph embedding, Knowl.-Based Syst. 313 (2025) 113040.

[21] B. Perozzi, R. Al-Rfou, S. Skiena, Deepwalk: Online learning of social representations, in: Proceedings of the 20th ACM SIGKDD International Conference on Knowledge Discovery and Data Mining, 2014, pp. 701–710.

[22] A. Grover, J. Leskovec, Node2vec: Scalable feature learning for networks, in: Proceedings of the 22nd ACM SIGKDD International Conference on Knowledge Discovery and Data Mining, 2016, pp. 855–864.

[23] W.L. Hamilton, Z. Ying, J. Leskovec, Inductive representation learning on large graphs, in: I. Guyon, U. von Luxburg, S. Bengio, H.M. Wallach, R. Fergus, S.V.N. Vishwanathan, R. Garnett (Eds.), Advances in Neural Information Processing Systems 30: Annual Conference on Neural Information Processing Systems 2017, December 4-9 2017, Long Beach, CA, USA, 2017, pp. 1024–1034.

[24] Z. Hu, Y. Dong, K. Wang, Y. Sun, Heterogeneous graph transformer, in: Proceedings of the Web Conference 2020, 2020, pp. 2704–2710.

[25] J. Xiao, Q. Dai, X. Xie, J. Lam, K. Kwok, Adversarially regularized graph attention networks for inductive learning on partially labeled graphs, Knowl.-Based Syst. 268 (2023) 110456, http://dx.doi.org/10.1016/J.KNOSYS.2023.110456.

[26] H. Peng, R. Zhang, Y. Dou, R. Yang, J. Zhang, P.S. Yu, Reinforced neighborhood selection guided multi-relational graph neural networks, ACM Trans. Inf. Syst. (TOIS) 40 (4) (2021) 1–46.

[27] P. Velickovic, G. Cucurull, A. Casanova, A. Romero, P. Liò, Y. Bengio, Graph attention networks, in: 6th International Conference on Learning Representations, ICLR 2018, Vancouver, BC, Canada, April 30 - May 3 2018, Conference Track Proceedings, OpenReview.net, 2018.

[28] A. Vaswani, N. Shazeer, N. Parmar, J. Uszkoreit, L. Jones, A.N. Gomez, L. Kaiser, I. Polosukhin, Attention is all you need, in: I. Guyon, U. von Luxburg, S. Bengio, H.M. Wallach, R. Fergus, S.V.N. Vishwanathan, R. Garnett (Eds.), Advances in Neural Information Processing Systems 30: Annual Conference on Neural Information Processing Systems 2017, December 4-9 2017, Long Beach, CA, USA, 2017, pp. 5998–6008.

[29] J. Baek, D.B. Lee, S.J. Hwang, Learning to extrapolate knowledge: Transductive few-shot out-of-graph link prediction, in: H. Larochelle, M. Ranzato, R. Hadsell, M. Balcan, H. Lin (Eds.), Advances in Neural Information Processing Systems 33: Annual Conference on Neural Information Processing Systems 2020, NeurIPS 2020, December 6-12 2020, virtual, 2020.

[30] L. Hu, A two-step method for classifying political partisanship using deep learning models, Soc. Sci. Comput. Rev. 42 (4) (2024) 961–976.

[31] Z. Wu, S. Pan, F. Chen, G. Long, C. Zhang, P.S. Yu, A comprehensive survey on graph neural networks, IEEE Trans. Neural Netw. Learn. Syst. 32 (1) (2021) 4–24, http://dx.doi.org/10.1109/TNNLS.2020.2978386.

[32] H.I. Aslan, C. Choi, H. Ko, Classification of vertices on social networks by multiple approaches, 2023, http://dx.doi.org/10.48550/ARXIV.2301.11288, CoRR abs/2301.11288. arXiv:2301.11288.

[33] L. He, C. Lu, J. Ma, J. Cao, L. Shen, P.S. Yu, Joint community and structural hole spanner detection via harmonic modularity, in: B. Krishnapuram, M. Shah, A.J. Smola, C.C. Aggarwal, D. Shen, R. Rastogi (Eds.), Proceedings of the 22nd ACM SIGKDD International Conference on Knowledge Discovery and Data Mining, San Francisco, CA, USA, August 13-17, 2016, ACM, 2016, pp. 875–884, http://dx.doi.org/10.1145/2939672.2939807.

[34] J. Zhou, G. Cui, S. Hu, Z. Zhang, C. Yang, Z. Liu, L. Wang, C. Li, M. Sun, Graph neural networks: A review of methods and applications, AI Open 1 (2020) 57–81, http://dx.doi.org/10.1016/J.AIOPEN.2021.01.001.

[35] R.C. Souza, R.M. Assunção, D.M. Oliveira, D.B. Neill, W. Meira Jr., Where did i get dengue? detecting spatial clusters of infection risk with social network data, Spat. Spatio-Temporal Epidemiology 29 (2019) 163–175.

[36] Q. Gong, Y. Liu, J. Zhang, Y. Chen, Q. Li, Y. Xiao, X. Wang, P. Hui, Detecting malicious accounts in online developer communities using deep learning, IEEE Trans. Knowl. Data Eng. 35 (10) (2023) 10633–10649, http://dx.doi.org/10.1109/TKDE.2023.3237838.

[37] J. Wang, X. He, Q. Gong, Y. Chen, T. Wang, X. Wang, Deep learning based malicious account detection in the momo social network, in: 2018 27th International Conference on Computer Communication and Networks, ICCCN, IEEE, 2018, pp. 1–2.

[38] M. Petkovic, M. Ceci, G. Pio, B. Skrlj, K. Kersting, S. Dzeroski, Relational tree ensembles and feature rankings, Knowl.-Based Syst. 251 (2022) 109254, http://dx.doi.org/10.1016/J.KNOSYS.2022.109254.

[39] M. Ceci, A. Appice, D. Malerba, Mr-sbc: A multi-relational naïve bayes classifier, in: N. Lavrac, D. Gamberger, H. Blockeel, L. Todorovski (Eds.), Knowledge Discovery in Databases: PKDD 2003 7th European Conference on Principles and Practice of Knowledge Discovery in Databases, Cavtat-Dubrovnik, Croatia, September 22-26 2003, Proceedings, in: Lecture Notes in Computer Science, vol. 2838, Springer, 2003, pp. 95–106.

[40] T. Mikolov, K. Chen, G. Corrado, J. Dean, Efficient estimation of word representations in vector space, in: Y. Bengio, Y. LeCun (Eds.), 1st International Conference on Learning Representations, ICLR 2013, Scottsdale, Arizona, USA, May (2013) 2-4, Workshop Track Proceedings, 2013.

[41] T. Mikolov, I. Sutskever, K. Chen, G.S. Corrado, J. Dean, Distributed representations of words and phrases and their compositionality, Adv. Neural Inf. Process. Syst. 26 (2013).

[42] G.D. Martino, G. Pio, M. Ceci, PRILJ: An efficient two-step method based on embedding and clustering for the identification of regularities in legal case judgments, Artif. Intell. Law 30 (3) (2022) 359–390, http://dx.doi.org/10.1007/S10506-021-09297-1.

[43] G. De Martino, G. Pio, M. Ceci, Multi-view overlapping clustering for the identification of the subject matter of legal judgments, Inform. Sci. 638 (2023) 118956.

[44] D.E. Rumelhart, J.L. McClelland, P.R. Group, et al., Parallel Distributed Processing, Foundations 1, 1988.

[45] C. Bellinger, S. Sharma, N. Japkowicz, One-class versus binary classification: Which and when? in: 2012 11th International Conference on Machine Learning and Applications, vol. 2, 2012, pp. 102–106, http://dx.doi.org/10.1109/ICMLA.2012.212.

[46] B. Krawczyk, Learning from imbalanced data: open challenges and future directions, Prog. Artif. Intell. 5 (4) (2016) 221–232.

[47] T.-Y. Lin, P. Goyal, R. Girshick, K. He, P. Dollár, Focal loss for dense object detection, in: Proceedings of the IEEE International Conference on Computer Vision, 2017, pp. 2980–2988.

[48] S. Nazir, M. Asif, S.A. Sahi, S. Ahmad, Y.Y. Ghadi, M.H. Aziz, Toward the development of large-scale word embedding for low-resourced language, IEEE Access 10 (2022) 54091–54097.

[49] M. McVicar, B. Di Giorgi, B. Dundar, M. Mauch, Lyric document embeddings for music tagging, 2021, arXiv preprint arXiv:2112.11436.

[50] L. Dong, N. Yang, W. Wang, F. Wei, X. Liu, Y. Wang, J. Gao, M. Zhou, H.-W. Hon, Unified language model pre-training for natural language understanding and generation, Adv. Neural Inf. Process. Syst. 32 (2019).

[51] Z. Chen, H. Mao, H. Li, W. Jin, H. Wen, X. Wei, S. Wang, D. Yin, W. Fan, H. Liu, et al., Exploring the potential of large language models (llms) in learning on graphs, ACM SIGKDD Explor. Newsl. 25 (2) (2024) 42–61.

[52] Z. Yang, Z. Dai, Y. Yang, J. Carbonell, R.R. Salakhutdinov, Q.V. Le, Xlnet: Generalized autoregressive pretraining for language understanding, Adv. Neural Inf. Process. Syst. 32 (2019).

[53] L.v.d. Maaten, G. Hinton, Visualizing data using t-sne, J. Mach. Learn. Res. 9 (2008) 2579–2605.

[54] M.T. Ribeiro, S. Singh, C. Guestrin, Why should i trust you? explaining the predictions of any classifier, in: Proceedings of the 22nd ACM SIGKDD International Conference on Knowledge Discovery and Data Mining, 2016, pp. 1135–1144.